\documentstyle[aaspptwo,rotate]{article}

\def\swan{\hbox{Swan C$_2$}}
\let\simgt\gtrsim
\let\simlt\lesssim
\makeatletter
\def\thebibliography{\subsection*{REFERENCES}
\list{}{\labelwidth3em\leftmargin\labelwidth\labelsep\z@\parsep\z@
\itemsep\z@\itemindent-3em\usecounter{enumi}}
\def\refpar{\relax}
\def\newblock{\hskip .11em plus .33em minus .07em}
\sloppy\clubpenalty4000\widowpenalty4000
\sfcode`\.=1000\relax}
\makeatother
\begin{document}
\title{Spectroscopy of the white-dwarf companions of PSR~0655+64 and
0820+02\footnote{Based on observations obtained at the W.~M.~Keck
Observatory on Mauna Kea, Hawaii, which is operated jointly by the
California Institute of Technology and the University of
California.}$^{,}$\footnote{This is a preprint of a paper accepted by
{\em The Astrophysical Journal (Letters)}. No bibliographic reference
should be made to this preprint.  Permission to cite material in this
paper must be received from the authors.}}
\author{M. H. van Kerkwijk, S. R. Kulkarni}
\affil{Palomar Observatory, California Institute of Technology 105-24,
       Pasadena, CA 91125, USA}
\begin{abstract}
We present spectra of the white-dwarf companions of the radio pulsars
0655+64 and 0820+02.  For the latter, we find a spectrum showing
strong lines of hydrogen, i.e., that of a DA star.  From modelling
these lines, the mass of a white dwarf can, in principle, be
determined accurately, thus leading to constraints on the evolution of
the binary and the mass of the neutron star.  Our present spectrum is
not of sufficient quality to set a strong limit, but it does indicate
that the white dwarf most likely has a low mass.  This is consistent
with the star being a helium white dwarf, as would be expected from
considerations of the mass function and the preceding evolution.  From
similar considerations, the companion of PSR~0655+64 is expected to be
a more massive, carbon-oxygen white dwarf.  This is confirmed by our
spectra, which show the Swan bands of molecular carbon, making it a DQ
star.  Unlike what is observed in other DQ stars, the strength of the
\swan\ bands changes drastically, by a factor two in about two hours.
We suggests this reflects large-scale surface inhomogeneities which
are rotated in and out of the observed hemisphere.  If so, this would
imply that the white dwarf rotates supersynchronously.
\end{abstract}


\section{Introduction}

About one tenth of the known radio pulsars reside in a binary system.
Two have massive, early-type companions, and have properties rather
similar to the average isolated radio pulsar.  These presumably have
been formed in a supernova explosion of a star in an early-type
binary.  Most of the others have evolved companions --- either white
dwarfs or neutron stars --- and differ markedly from (most of) their
isolated counterparts, showing more rapid spin periods and smaller
magnetic fields (for recent reviews, see Bhattacharya \& Van den
Heuvel \cite{bhatvdh:91}; Verbunt \cite{verb:93}; Phinney \& Kulkarni
\cite{phink:94}; Kulkarni \cite{kulk:95}; these references have been
used throughout this section).  The rapid spin is generally believed
to result from a phase of mass transfer in the evolutionary history of
the binary, during which a large amount of mass and angular momentum
is accreted.  The reduction of the magnetic field is thought to occur
in this phase as well, but the physical mechanism underlying this
(e.g., it being ``buried'' by the accreted matter) is not understood.

The radio pulsars with evolved companions can be divided into three
groups, generally referred to as the low-, intermediate- and high-mass
binary pulsars (hereafter, LMBP, IMBP and HMBP, respectively).  While
these classifications are made on the basis of the inferred mass of
the companion (from the mass function), stellar evolutionary scenarios
allow us to relate these systems to the descendents of binaries
composed of a neutron star and a low-mass ($M\simlt1\,M_\odot$),
intermediate-mass ($1\,M_\odot\simlt M\simlt8\,M_\odot$) and high-mass
($M\simgt8\,M_\odot$) secondary, respectively.  The expectation is
that these secondaries evolve to Helium white dwarfs, carbon-oxygen
(C-O) white dwarfs, and neutron stars, respectively.  There is a fairly
systematic trend of decrease in magnetic field strength as one
proceeds from HMBPs to LMBPs.  Presumably, this reflects the decrease
in speed with which evolution proceeds in these systems, and the
corresponding increase of the total amount of matter that is accreted.

Observations of the white-dwarf companions provide a number of
diagnostics for these objects.  For instance, the surface temperature
of the white dwarf can be used to infer a cooling age, which sets a
lower limit to the age of the neutron star.  Such limits have been
derived from broad-band photometry for PSR~0655+64 (Kulkarni
\cite{kulk:86}), 0820+02 (ibid.; Koester, Chanmugan, \& Reimers
\cite{koescr:92}), 1855+09 (Callanan et al.\ \cite{call&a:89};
Kulkarni, Djorgovski, \& Klemola \cite{kulkdk:91}), J0437$-$4715
(Bailyn \cite{bail:93}; Bell, Bailes, \& Bessel \cite{bellbb:93};
Danziger, Baade, \& Della Valle \cite{danzbdv:93}), and J1012+5307
(Lorimer et al.\ \cite{lori&a:95}), and they provide the strongest
evidence so far that magnetic fields of neutron stars do not decay on
a time scale of millions of years, as had been thought before.  In the
absence of good constraints on the masses of the white dwarfs,
however, the cooling ages cannot be very accurately determined.

Mass determinations would not only lead to better constraints on the
cooling ages, but also allow one to verify, e.g., the orbital-period,
white-dwarf mass relation predicted for the LMBP (Refsdal \& Weigert
\cite{refsw:71}; Savonije \cite{savo:87}; Joss, Rappaport, \& Lewis
\cite{jossrl:87}; Rappaport et al.\ \cite{rapp&a:95}), or to constrain
the mass of the neutron star.  In addition, they could be used to
obtain independent distance estimates to the systems, which can be
compared to those derived from the dispersion measure of the pulsar.

Constraints on the white-dwarf mass can be derived from Shapiro delay,
but this is possible for only a few binaries with favorable
geometries.  Another possibility is to use spectroscopy of the white
dwarfs.  In recent years, much progress has been made in the
spectroscopic determination of surface gravities (and hence masses and
radii), especially for white dwarfs of spectral type DA, i.e., those
with a hydrogen atmosphere (e.g., Bergeron, Saffer, \& Liebert
\cite{bergsl:92}).  Spectroscopy of the very faint companions of
binary pulsars is non-trivial, and has so far only been attempted for
the brightest, that of PSR~J0437$-$4715 (Danziger et al.\
\cite{danzbdv:93}).  Unfortunately, that white dwarf is rather cool
($T\simeq4000\,$K), and no line features were detected.

Here, we present spectroscopy of two somewhat fainter, but hotter
companions, one of an IMBP, PSR~0655+64, and one of a LMBP,
PSR~0820+02.  For recent radio studies of these binaries, see Taylor
\& Dewey (\cite{tayld:88}) and Jones \& Lyne (\cite{jonel:88}).  For
previous optical studies, see the references listed above.

\section{Observations}

Spectra were taken on New Year's eve of 1995 at the Keck 10\,m
telescope with the Low-Resolution Imaging Spectrometer (LRIS).  With
the $300\,{\rm{}line\,mm}^{-1}$ grating, the wavelength range of 3750
to 8780\,\AA\ was covered at at $2.5\,{\rm\AA}\,{\rm{}pix}^{-1}$.  The
observing conditions were good throughout the night, but a substantial
amount of time was lost due to telescope problems.

Three spectra were taken of the companion of PSR~0655+64
($V=22.2\,$mag), starting at 1 January 1995, 8:54, 9:56 and
10:16~{\sc{}ut}, with integration times of 30, 15 and 45 minutes,
respectively.  For the first spectrum, the slit was positioned over
the companion and a nearby elliptical galaxy (see Kulkarni
\cite{kulk:86}), while for the latter two it was set close to the
parallactic angle.  A 0.7\arcsec\ wide slit was used, giving a
resolution of $\sim\!8\,$\AA.  The companion of PSR~0820+02
($V=22.8\,$mag) was observed for 45 minutes starting at
14:26~{\sc{}ut}, using a 1\arcsec\ slit to improve the throughput
(leading to a resolution of $\sim\!12\,$\AA).  For approximate flux
calibration, the spectrophotometric standard HD\,84937 (Oke \& Gunn
\cite{okeg:83}) was observed.

\begin{figure*}[t]
{\centering\leavevmode%
\rotate[r]{\epsfysize\hsize\epsfbox{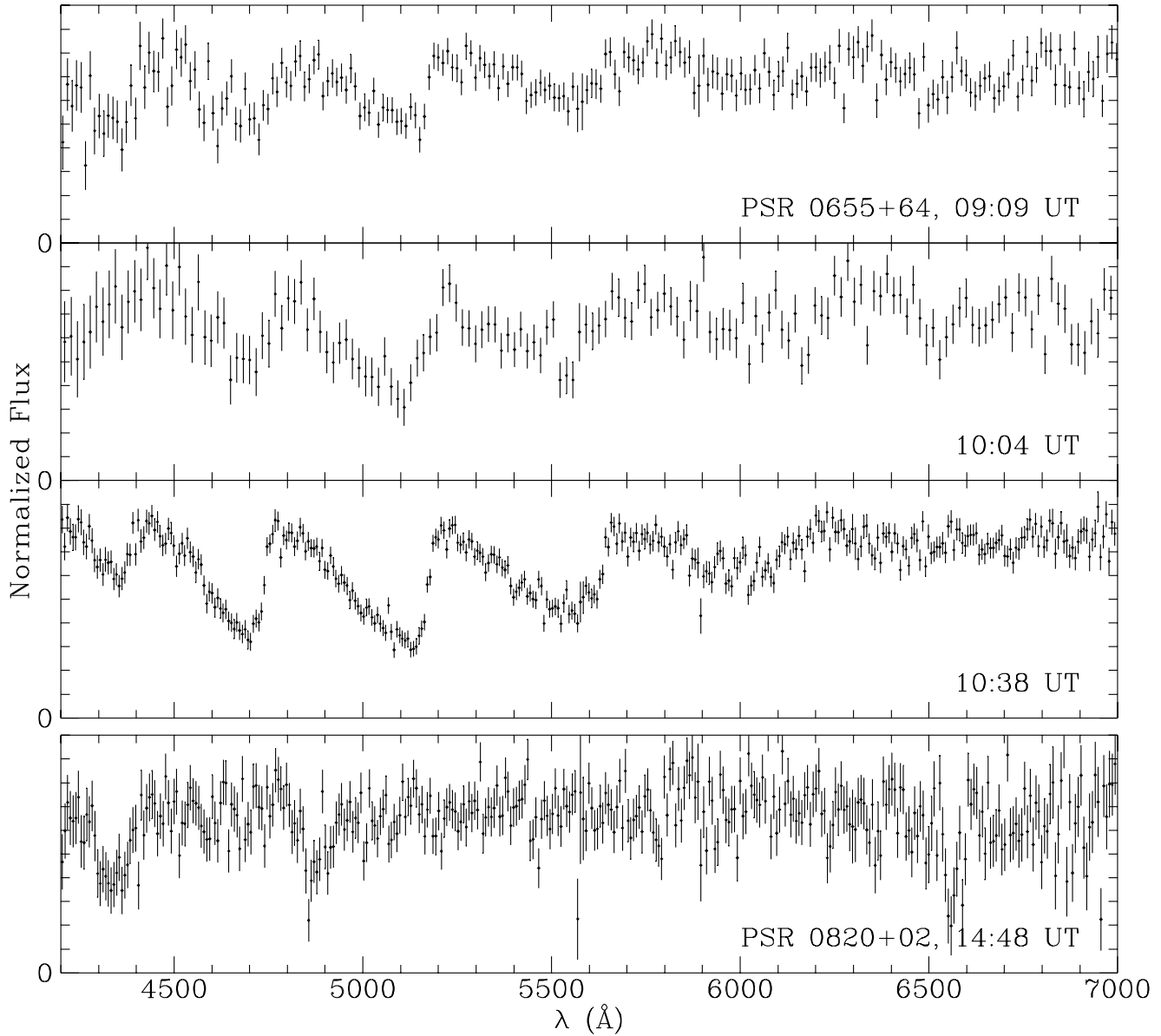}}}
\caption{The spectra of the companions of PSR~0655+64 (top three
panels) and PSR~0820+02 (bottom panel).  The spectra of the former all
show \swan\ bands in absorption, but with varying strength, while the
spectrum of the latter shows absorption lines of H$\alpha$, H$\beta$,
and H$\gamma$.  For each spectrum, the mid-exposure time is indicated.
For the representation shown here, the spectra were binned in (from
top to bottom) 5-, 7-, 3-, and 3-pixel wide bins, and the rather noisy
parts shortward of 4200\,\AA\ and longward of 7000\,\AA\ were
omitted.\label{fig:spectra}}
\end{figure*}

The reduction of all spectra was done using MIDAS\footnote{The Munich
Image Data Analysis System is developed and maintained by the European
Southern Observatory.} and programs running in the MIDAS environment.
The frames were bias-corrected, flat-fielded and sky-subtracted using
standard procedures.  The spectra were extracted using an
optimal-extraction method similar to that presented by Horne
(\cite{horn:86}).  It turned out that the different exposures of the
flux standard were inconsistent in both level and slope of the
continuum.  Since flux calibration was thus impossible, we normalized
the spectra for the representation shown here
(Fig.~\ref{fig:spectra}).  This was done by dividing by a continuum
defined by one of the flux-standard spectra, and scaling by
$\lambda^\beta$, with $\beta$ chosen such that the continuum appeared
straight.

\section{Results and Discussion}

\subsection{PSR~0655+64}

The spectra of the companion of PSR~0655+64 (Fig.~\ref{fig:spectra})
are those of a DQ star, showing strong \swan\ bands (for a review of
white-dwarf spectra, see Wesemael et al.\ \cite{wese&a:93}).  The
\swan\ bands are thought to be due to traces of carbon in a
helium-rich atmosphere, brought up by convection from a deeper region,
which is enriched in carbon due to upwards diffusion of carbon from
the core (Pelletier et al.\ \cite{pell&a:86}).  Thus, the presence of
the Swan bands directly confirms the expectation that the object is a
carbon-oxygen white dwarf.

The presence of the Swan bands, combined with the absence of lines
from atomic carbon, also indicates that the temperature is between
6000 and 9000~K (inferred from Wegner \& Yackovich \cite{wegny:84} and
Wesemael et al.\ \cite{wese&a:93}).  This compares well with the range
of 5500 to 8000~K derived from photometry of this star (Kulkarni
\cite{kulk:86}), and hence provides independent confirmation of the
conclusion of Kulkarni (\cite{kulk:86}) that the cooling age of the
white dwarf is about $2\times10^9\,$yr, comparable to the
characteristic age of $P/2\dot{P}=3.6\times10^9\,$yr of the pulsar.

{}From the three spectra that we have, one can see that the strength of
the C$_2$ features is variable.  In fact, the total equivalent width
changes by a factor of 2 over the two hours spanned by the
observations, from about 170 to 330$\,$\AA\ in the range
4430--5650\AA.  As far as we know, this is unprecedented.  Since white
dwarfs in general are not particularly variable stars, it is tempting
to relate the changes to fixed surface patterns --- due to, e.g.,
abundance or temperature variations, perhaps related to the presence
of a magnetic field as in magnetic Ap stars (Borra, Landstreet, \&
Mestel \cite{borrlm:82}) --- that are rotated in and out of the
observed hemisphere.  If so, then from the swiftness of the change in
the spectrum, the strength of the features, and the integration times,
it is easy to see that a possible periodicity cannot be shorter than
about 3 hours (otherwise, the changes would be washed out), or longer
than about 12 hours (to allow a change from about 30 to about 70\%
depth of the strongest part of the absorption).  This is substantially
shorter than the \hbox{$\sim\!1$-day} orbital period, and thus an
association with orbital variations --- such as could be produced by,
e.g., heating of one hemisphere --- seems unlikely.

If we associate the variations with the rotation of the white dwarf,
then its rotation period has to be two to eight times shorter than the
orbital one.  This might actually not be unexpected: the progenitor of
the white dwarf was a helium giant transferring mass to the pulsar
(e.g., Iben \& Tutukov \cite{ibent:93}), and if it was rotating
synchronously --- as does not seem implausible given the extremely
circular orbit ($e=7.5\,10^{-6}$; Jones \& Lyne \cite{jonel:88}) ---
then it would have been spun up due to conservation of angular
momentum when it shrunk to form a white dwarf.  In fact, we can set a
rough upper limit to the mass of the envelope that fell back onto the
white dwarf if we assume that the effect due to the shrinking of the
core can be neglected.  For this case, the moment of inertia in the
envelope has to be at least equal that of the core -- or equivalently
(by assumption) that of the white dwarf -- in order to be able to spin
it up by at least a factor two.  Hence, the mass of the envelope
should be $\ga{}M_{\rm{}WD}R_{\rm{}WD}^2/R_{\rm{}env}^2$ (ignoring
differences in structure).  Since the progenitor was filling its Roche
lobe, $R_{\rm{}env}\simeq R_{\rm{}L}
\simeq(GM_{\rm{}WD}/10\Omega_{\rm{}orb}^2)^{1/3} \simeq 2\,R_\odot$
(using Eq.~10.1 of Phinney \& Kulkarni \cite{phink:94}).  With
$M_{\rm{}WD}\simeq0.8\,M_\odot$ and $R_{\rm{}WD}\simeq0.01\,R_\odot$,
one finds a mass of the envelope of a few $10^{-4}\,M_\odot$.
Interestingly, this is similar to the mass of the helium envelope that
is inferred for DQ stars from the presence and strength of the carbon
features (Pelletier et al.\ \cite{pell&a:86}; but see also Weideman \&
Koester \cite{weidk:95}; Dehner \& Kawaler \cite{dehnk:95}).

While we realise that our estimate is a rough one -- the very
existence of a periodicity still needs to be confirmed -- we note
that, in principle, it might be possible to use rotation periods of
white dwarfs in similar systems to constrain the final phases of the
evolution.  Especially for the systems with somewhat longer orbital
periods, where the angular momentum is dominated by the envelope, the
final spin rate should depend almost uniquely on the radius of the
giant.  Since the latter is related to the orbital period, one might
expect, e.g., to find a correlation between the spin period and
orbital period in such systems.

\subsection{PSR~0820+02}

In contrast to the companion of PSR~0655+64, this white dwarf shows
strong lines of hydrogen (see Fig.~\ref{fig:spectra}), making it a DA
white dwarf.  As mentioned in the introduction, for a DA white dwarf
it is possible to determine surface temperature and gravity uniquely
from the spectrum.  From a first comparison of the spectrum with model
atmospheres, kindly done for us by Dr.~Bergeron, it follows that it is
consistent with the temperature being in the range of 14000 to 16500~K
found by Koester et al.\ (\cite{koescr:92}).  The best-fitting surface
gravities for that range would indicate a mass of 0.25 to
0.35\,$M_\odot$, consistent with the idea that it is a helium white
dwarf.  However, the spectrum is of insufficient quality to set a
strong limit: the 95\% upper limit to the mass is
$\sim\!0.9\,M_\odot$.  Thus, it is not yet possible to verify whether
the mass is within the range of 0.42 to 0.60$\,M_\odot$ expected from
the orbital-period, white-dwarf mass relation (Rappaport et al.\
\cite{rapp&a:95}).

{}From photometry, combined with an upper limit of 1.9\,kpc to the
distance, Koester et al.\ (\cite{koescr:92}) derived a lower limit of
about $0.5\,M_\odot$ to the mass of the companion of PSR~0820+02.
Indeed, if the white dwarf were to have a mass of about
$0.3\,M_\odot$, it would have $M_V\simeq10.4\,$mag (Bergeron, private
communication), and with $V\simeq22.8\,$mag and $A_V\simeq0.1\,$mag
(Koester et al.\ \cite{koescr:92}), the system would be at a distance
of $\sim\!2.8\,$kpc.  We note, however, that the distance limit
Koester et al.\ used was derived by taking twice the distance of
0.95\,kpc indicated by the dispersion measure of
$23.6(2)\,{\rm{}cm^{-3}\,pc}$ (Taylor \& Dewey \cite{tayld:88})
combined with the model of the Galactic electron distribution of Lyne,
Manchester, \& Taylor (\cite{lynemt:85}).  With the more recent model
of Taylor \& Cordes (\cite{taylc:93}), one would infer a distance of
1.4\,kpc.  Taylor \& Cordes estimate that the mean uncertainty in the
new dispersion-measure derived distances is about 25\%, but they note
that there is a dependence on the position of the pulsar.  For
PSR~0820+02, at $l^{\rm{}II}=222^\circ$, $b^{\rm{}II}=21^\circ$, we
find from their Figures~7 and 8 that the model is substantially more
uncertain.

While it is thus at present not possible to constrain the mass
unambiguously, we note that for a $0.6\,M_\odot$ white dwarf, Koester
et al.\ (\cite{koescr:92}) derived that the cooling age was within an
``allowed range'' of 1.5 to $2.7\,10^8\,$yr, which seems only
marginally consistent with the characteristic age of $1.1\,10^8\,$yr
of the neutron star (which should be an upper limit to the true age).
If the white dwarf were to have a lower mass, the two age estimates
would be in better agreement.

\section{Conclusions}

We have presented spectra of the white-dwarf companions of the radio
pulsars 0655+64 and 0820+02.  From these spectra, combined with
published temperature determinations, we can classify the two as
DQ6--9 and DA3/4, respectively.  This confirms the expectation that
the former is a carbon-oxygen white dwarf, and is consistent with the
latter being a helium white dwarf.

As hoped, it will be possible to derive accurate temperatures and
surface gravities for the companion of PSR~0820+02.  While this will
likely be difficult for the companion of PSR~0655+64, its changing
spectrum may instead turn out to provide us with a unique possibility
to learn more about its previous evolution, as well as about the
formation of spectral features in white dwarfs.  It seems clear that
further spectral studies of both these and other pulsar companions are
warranted.

\acknowledgements We are very grateful to Pierre Bergeron for making
the model-atmosphere comparisons, and thank him, Brad Hansen, Yanqin
Wu, and Peter Goldreich for useful discussions.  M.H.v.K.\ is
supported by a NASA Hubble Fellowship and S.R.K.\ by grants from the
US NSF, NASA and the Packard Foundation.

\end{document}